\pdfoutput=1
\documentclass[10 pt, final, journal]{IEEEtran}
\IEEEoverridecommandlockouts	%
\usepackage{amsmath}
\usepackage{amssymb}
\usepackage{stmaryrd}
\usepackage[load-configurations=abbreviations]{siunitx}
\usepackage{hyperref}
\usepackage{graphicx}
\graphicspath{{images/}}
\usepackage{physics}
\usepackage{algorithm}%
\usepackage{algorithmic}
\usepackage{todonotes}
\usepackage[T1]{fontenc}
\usepackage[utf8]{inputenc}
\usepackage{tikz}
\usetikzlibrary{plotmarks}
\usepackage{tabularx}
\usepackage{subfiles}
\usepackage{caption}
\usepackage{subcaption}
\usepackage{float}
\usepackage{multirow}
\def\reals{\mathbb{R}}

\newcommand\copyrighttext{%
  \begin{footnotesize}
  © 2020 IEEE.  Personal use of this material is permitted.  Permission from IEEE must be obtained for all other uses, in any current or future media, including reprinting/republishing this material for advertising or promotional purposes, creating new collective works, for resale or redistribution to servers or lists, or reuse of any copyrighted component of this work in other works.
  \end{footnotesize}
}
\newcommand\copyrightnotice{%
\begin{tikzpicture}[remember picture,overlay]
\node[anchor=south,yshift=1pt] at (current page.south) {\fbox{\parbox{\dimexpr\textwidth-\fboxsep-\fboxrule\relax}{\copyrighttext}}};
\end{tikzpicture}%
}

\title{\LARGE \bf
	Validating feedback control to meet stiffness requirements in additive manufacturing
}

\author{K\'evin Garanger$^{1}$, Thanakorn Khamvilai$^{2}$, and Eric Feron$^{3}$%
	\thanks{$^{1,2,3}$All authors are with the Decision and Control Laboratory at Georgia Institute of Technology, Atlanta, GA, 30332, USA \{kevin.garanger, tkhamvilai3, feron\}@gatech.edu. Feron is on leave at King Abdullah University of Science and Technology, Thuwal, Saudi Arabia.}%
}
\date{}
\begin{document}
\maketitle
\copyrightnotice

\thispagestyle{empty}
\pagestyle{empty}

\begin{abstract}
	This paper discusses the possibility of making an object that precisely meets global structural requirements using additive manufacturing and feedback control. An experimental validation is presented
	by printing a cantilever beam with a prescribed stiffness requirement.
	The printing process is formalized as a model-based finite-horizon discrete control problem, where the control variables are the widths of the successive layers.
	Sensing is performed by making {\em in situ} intermediate stiffness measurements on the partially built part. The hypothesis that feedback control is effective in enabling the 3D-printed beam to meet precise stiffness requirements is validated experimentally.
\end{abstract}

\section{Introduction}\label{introduction}

	The possibilities offered by Additive Manufacturing (AM)~\cite{parthasarathy2011design} can be realized by understanding its ability to meet requirements related to part geometry, uniformity and thermal/structural properties, for example. Challenges to the achievement of this goal materialize by the lack of precise models of additively manufactured parts and an inherently noisy print process~\cite{bourell2009brief, frazier2014metal, huang2015additive}. AM systems, ranging from low-cost, fused filament printers to more complex metal printers, feature dozens or hundreds of parameters that may be tuned as the printing proceeds. 
	These parameters may include extrusion temperature, extrusion rate, print bed temperature, nozzle temperature, printing head path, etc. Tuning these parameters to improve part quality is often a time-consuming and empirical job that is not always successful.

    This paper considers the experimental validation of feedback control in AM to print a cantilever beam that must meet a global stiffness requirement.
    To the authors knowledge, this work is the first to directly consider a part's structural requirements to establish a feedback control strategy for the printing process.
    The consideration of structural properties during the manufacturing of a part is essential in applications with strict tolerances when the manufacturing process is subject to noise.
    For instance, airplane propeller blades are subject to tight requirements regarding their natural frequencies as to avoid vibrations that could lead to a loss of efficiency or structural failure.
    Manufacturing them using a closed-loop AM process could therefore benefit from including their structural requirements in the problem formulation.

	The remainder of this paper is structured as follows. Section~\ref{related_work} describes relevant prior work. In Section~\ref{cantilever}, the main experiment that constitutes the original contribution of this paper is described. The derivation of the closed-loop control architecture is given in Section~\ref{feedbacklaw}. The results of the printing experiments are finally presented in Section~\ref{experiments} before suggestions for further research and a conclusion are respectively given in Section~\ref{suggestions} and Section~\ref{conclusion}.

\section{Related work}\label{related_work}

	Two fields of research in AM are especially relevant to this paper. 
	
	First, there is modeling and characterizing the behavior of AM systems. %
    Because of the complexity and diversity of AM processes, the precise characterization of their behavior can only be done for a specific process and a specific material.
    For instance, reference \cite{mazumder1997direct} experimentally studies the effects of tempering temperature and laser traverse speed and power on the microstructure of a kind of steel, the H13 tool steel, manufactured with Direct Metal Deposition (DMD).
    The relation between a layer thermal history and its microhardness is evaluated for a similar process using low-alloy steel in~\cite{el2008phase}.
    In~\cite{thijs2010study}, different samples made of Ti-6Al-4V, a titanium alloy, are manufactured with selective laser melting by changing the laser scanning velocity, scanning strategy, and hatching space.
    The microstructure of these samples is then studied by light optical microscopy.
    References~\cite{zheng2008thermal} and~\cite{bontha2009effects} both model beam-based deposition methods to simulate the thermal behavior and solidification during the printing process. These simulations are used to study the influence of process parameters such as beam power, velocity, and diameter of the resulting microstructure.
    The thesis~\cite{antonysamy2012microstructure} models the thermal behavior of Ti-6Al-4V for three different AM processes. For one of them, which is wire arc additive manufacturing, the influence of process parameters on the grain structure is studied through simulation and experimental validation.
    Experimental work with Fused Deposition Modeling (FDM) includes~\cite{ahn2002anisotropic} and~\cite{bagsik2010fdm}, which both study the compressive and tensile stengths of samples printed with different parameters.
    In~\cite{ahn2002anisotropic}, the parameters of interest are air gap, bead width, model temperature, ABS color, and raster orientation, while in~\cite{bagsik2010fdm}, the influence of the parts orientation is studied.
    Some works about the modeling of AM processes such as~\cite{sammons2019two} and \cite{guo2018control} focus on defining and identifying deposition models suitable for closed-loop control strategies aiming at improving the geometrical accuracy of printed parts.

	Second, there is the optimization of the printing parameters and the control of the printing processes via feedback to improve the properties of the manufactured parts.
    The concept of layer-to-layer control was introduced in~\cite{tang2011layer}. In layer-to-layer control, measurements typically including height sensing are made after a layer is deposited and used to adapt the process parameters for the printing of the next layer in order to achieve a specific height profile.
    The feedback is therefore performing sequentially only after a new layer is printed completely. This is due to the challenges of sensing AM processes during the printing of a layer.
    The work in~\cite{tang2011layer} is based on a model of the laser metal deposition process derived using mass, energy, and momentum conservation. It uses Iterative Learning Control to adapt the powder flow rate in order to track the reference height profile.
    More recent works have used layer-to-layer control applied to different AM processes and using different control strategies with the goal of achieving a prescribed part geometry. The work of~\cite{lu2014layer} introduces a height change model for ink-jet 3D printing based on the droplet pattern. This model is then used in conjunction with Model Predictive Control (MPC)~\cite{mpc} to minimize the error with a reference height at each layer.
    In~\cite{guo2017distributed}, a similar approach is taken, except that a graph-based height evolution model is used with a distributed MPC algorithm, reducing the computation time significantly.
    The work by~\cite{xiong2016closed} relies on a PID controller to adjust the travel speed in wire and arc additive manufacturing to track a reference layer width.
    The control of the freeze-form extrusion process is done in~\cite{zomorodi2016extrusion} with a high-level controller prioritizing between two objectives related to the flow of the paste and a low-level controller based on universal function approximator trained with MPC to satisfy the objective of the top controller.
    In \cite{sammons2018repetitive}, a laser metal deposition process if formulated as a repetitive process control problem and a PI controller is used to adjust a set of process variables in order to track a reference height profile.
    The work presented in~\cite{nassar2015intra} is one of the few using intra-layer control feedback of the build path for the control of microstructure in directed energy AM.

    In summary, the existing work regarding the use of feedback control of AM processes focuses on controlling geometrical properties or local microstructures of the part being manufactured.
    The matter of interest to this paper contrasts with the foregoing works as follows.
\begin{itemize}
\item {\bf Geometrical vs. mechanical properties:} All the papers mentioned previously that use layer-to-layer control track a reference height profile during the printing process. Instead of controlling a part's geometric features, we focus on its mechanical properties.
\item {\bf Planning horizon:} Our process relies on a macroscopic model comprised of only one control and one state variable per layer. This makes the trajectory optimization over the complete remaining printing process tractable and removes the necessity to regulate around a nominal trajectory or to plan the trajectory on a receding horizon, as done in the layer-to-layer works described in the foregoing.
\end{itemize}
Supporting these choices and the pertinence of this paper are the difficulty to reliably predict macroscopic structural properties of a printed part based on its geometry alone, on the one hand~\cite{ahn2002anisotropic}, and the value of identifying and meeting new performance metrics for printed parts, on the other hand. 
The problem of printing a part to meet specific stiffness requirements builds upon our prior work~\cite{garanger20183d}, where a leaf spring is manufactured using AM, validating the hypothesis that introducing a high level feedback control loop during the 3D printing of an object enables it to meet stiffness requirements more precisely. This paper builds upon~\cite{garanger20183d} by showing that such requirements can be met via feedback control during the printing of a single part with \emph{in situ}
sensing.

\section{Manufacturing a cantilever beam to meet stiffness specifications}\label{cantilever}

	\begin{figure}
		\centering
		\framebox{\parbox{.4\textwidth}{\includegraphics[width=.4\textwidth]{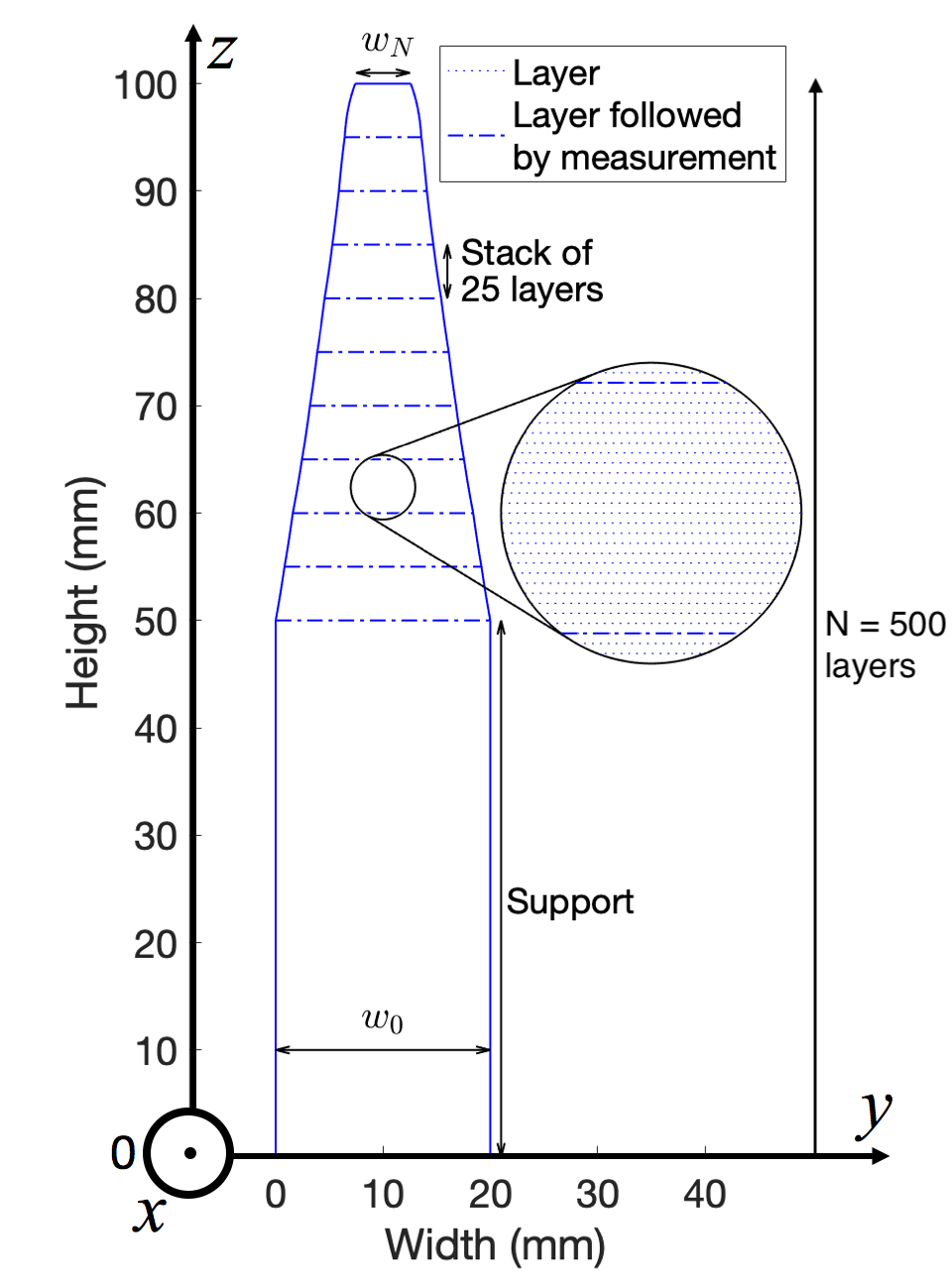}}}
		\caption{General characteristics of printed specimen. Individual layers are 0.2 mm thick, and there are $N=500$ layers total. Under closed loop operation, measurements are performed once every $25$ layers (\SI{5}{\milli\metre}). Measurements locations are shown by dash-dotted lines.}
		\label{img:shape_diagram}
	\end{figure}

	\subsection{Hypothesis and experiment overview}
	The hypothesis examined in this paper is whether feedback control enables a relatively low-grade printer to produce parts that
	 meet stiffness requirements more precisely than otherwise.
	The experiment used to validate this hypothesis consists of printing a cantilever beam, such as that shown in Fig.~\ref{img:shape_diagram}. A rudimentary 3D printer, described in the experiment section thereafter, is used with model-based feedback control for the lateral stiffness of the beam to reach a specific value. At any time during the printing process, the stiffness is defined as the limit of the ratio of the load to the deflection of a beam as the load tends towards 0. In practice, small finite loads are applied and geometric nonlinear effects are neglected.

The control variable 
is the width of the beam, which can be adjusted, under constraints discussed below, for each layer.
The sensory signal is the transverse stiffness of the unfinished beam measured at its tip.  
Measurements are performed periodically. Based on current and past measurements, a sequence of future layers is designed and executed by a dedicated control procedure. The process iterates until the print job is completed. Details are now provided.
%
%

\iffalse	
	 \begin{itemize}
	 	\item Printing of the part foundation ($\approx$ one hour)
	 	\item Control loop ($\approx$ \si{3.5} hours):
	 		\item Stiffness measurement ($\approx$  \si{30} seconds)
	 		\item Control law computation ($\approx$  \si{30} seconds)
	 		\item Printing of a fixed number of layers ($\approx$  \si{20} minutes)
	 \end{itemize}
	
	\begin{algorithmic}
		\renewcommand{\algorithmicrequire}{\textbf{Input:}}
		\renewcommand{\algorithmicensure}{\textbf{Output:}}
		\REQUIRE $N$ (number of layers), $K$ (target stiffness), $S$ (measurement schedule)
%
%
%
%
		\FOR {$n = 1$ to $N$}
		\IF {measurement scheduled at $n$ in $S$}
		\STATE Perform stiffness measurement
		\STATE Update subsequent layer widths to reach $K$
		\ENDIF
		\STATE Print one layer
		\ENDFOR
		\RETURN
	\end{algorithmic} 
\fi	
	\subsection{Printed beam model}

 A macroscopic probabilistic structural model of the beam as it is printed
 is derived and calibrated.
%
%
%
	\iffalse
	\begin{figure}
		\centering
		\framebox{\parbox{.3\textwidth}{\includegraphics[width=.3\textwidth]{open_loop_specimen_with_notations}}}
		\caption{Picture of a full specimen printed with an open-loop control.}
		\label{img:open_loop_specimen}
	\end{figure}
	\fi
	
	\subsubsection{Macroscopic structural model}

\begin{figure}[!htb]
		\centering
		\framebox{\parbox{.45\textwidth}{\includegraphics[width=.45\textwidth]{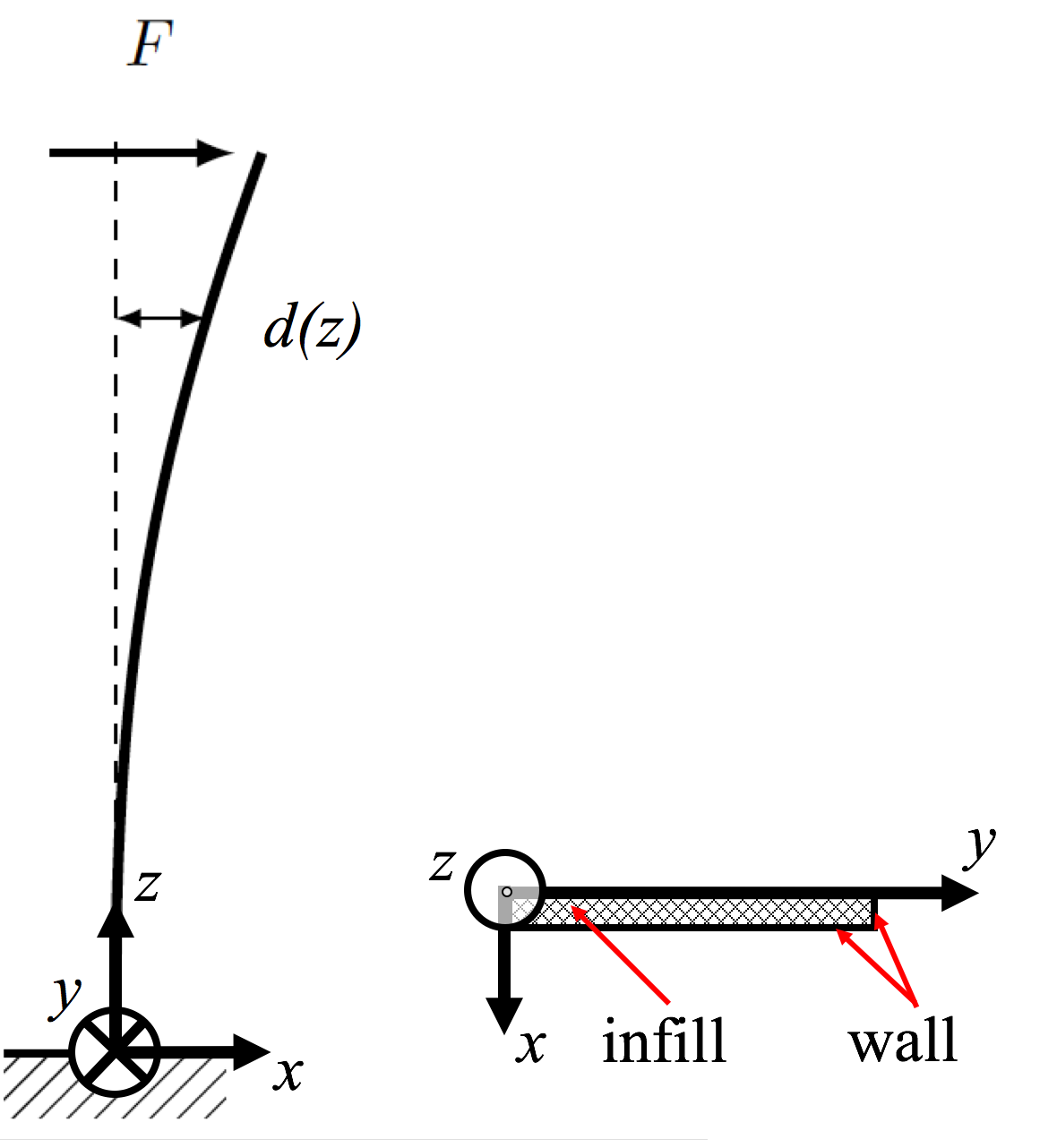}}}
		\caption{Beam deflection under load and details of beam cross-section}
		\label{deflection}	
	\end{figure}

Although objects made with FDM are known for their anisotropic properties \cite{ahn2002anisotropic}, we assume the relation between load and deflection in the cantilever beam deflection test satisfies the Euler-Bernoulli equation
	\begin{equation}
		\frac{\dd^2}{\dd z^2}\bigg(EI \frac{\dd^2d}{\dd z^2}\bigg) = 0,
		\label{e-b}
	\end{equation}
where $d=d(z)$ is the deflection (see Fig.~\ref{deflection}), $E$ is Young's modulus, and $I=I(z)$ is the second moment of area of the cross-section of the beam at height $z$ along the $x$ axis. For more details about the Euler-Bernoulli equation and the definition of the second moment of area, see~\cite{gere1997mechanics}. 
Because of the varying width of the printed beam, $I(z)$ is not constant. However, the beam can be divided in layers of equal height, with each single layer having a constant width. $I(z)$ is then a piecewise constant function.
For a cantilever beam made of $n$ layers of height $h$, let $I_k=I(kh)$ be the second moment of area of the $k$-th layer ($1\leq k \leq n$). Let $F$ be the applied horizontal load on top of the partially printed beam as shown in Fig.~\ref{deflection}. The solution to~(\ref{e-b}) is
	\begin{equation}\label{eq:deflection}
		d(nh) = \frac{h^3F}{E}\sum_{k=1}^{n}\frac{c_{n,k}}{I_k}
	\end{equation}
	with
	\begin{equation}\label{eq:coefficients}
		c_{n,k} = \frac{3(2k-1)(n-k)+3k^2-1}{6}.
	\end{equation}
Next, introduce $w_k = w(kh)$, the width of the beam for the $k$-th layer. We assume $I_k$ is affine in $w_k$ to capture the presence of two lateral walls printed at the end of each layer, see Fig.~\ref{deflection}, whose contribution to $I_k$ is constant, and that of the infill material and front and back walls, whose contribution grows linearly with $w_k$.
We therefore postulate the affine model
	\begin{equation}
		\label{eq:secondmom}
		\frac{E}{h^3F}I_k = \frac{1}{\alpha}(w_k+\gamma),
	\end{equation}
where $\alpha$ and $\gamma$ are parameters that capture the geometrical and material properties of the beam.

	\subsubsection{Dynamics of the printing process}
	
	Consider printing a beam with a total of $N$ layers seen as a dynamic process involving $N$ stages. 
Given $n\leq N$, introduce the state $s_n \in \reals^N$ vector of $N$ real values with $s_k = 1/w_k$ when $k\leq n$ and $0$ otherwise, with $s_0 = 0$.
The layer-to-layer print dynamics of the beam are given by the recursion
		\begin{equation}
			s_n = s_{n-1} + \frac{1}{u_n+\gamma+\epsilon_n}\boldsymbol{e_n},
		\label{eq:firstorder}
		\end{equation}
where $u_k$ is the desired layer width - the control variable - and $\epsilon_k$ is process noise such that $w_k = u_k+\epsilon_k$. $\boldsymbol e_n$  is the $n$-th canonical basis vector. We choose independent identically distributed normal distributions of mean $0$ and standard deviation $\sigma_p$ for the $\epsilon_k$. Under the assumption $\epsilon_n \ll \gamma$, we write
	\[
	s_n \simeq s_{n-1} + \frac{1}{u_n+\gamma}\left(1-\frac{\epsilon_n}{u_n+\gamma}\right)\boldsymbol{e_n}.
	\]
	Since $\epsilon_n$ has a symmetric p.d.f. about zero, we can also write
	\begin{equation}\label{eq:dynamics}
		s_n \simeq s_{n-1} + \frac{1}{u_n+\gamma}\left(1+\frac{\epsilon_n}{u_n+\gamma}\right)\boldsymbol{e_n}.
	\end{equation}
	
	\subsubsection{System output}
		The system's output is the compliance 
	\begin{equation}\label{eq:compliance}
		\mathcal{C} = \frac{d(nh)}{F} = \alpha\sum_{k=1}^{n}\frac{c_{n,k}}{w_k + \gamma} = \alpha C_n^Ts_n
	\end{equation}
	of the partially printed beam, where $C_n = (c_{n,1},\dots,c_{n,n})^T$. Let the measured quantity be a noisy stiffness reading 
	$c^k_n = \frac{1}{\mathcal{C}_n} + \nu_n$, where $\nu_n \sim N(0,\sigma_0)$  and $\sigma_o$ is given. The observed compliance at stage $n$ is
	\[
	o_n = \frac{1}{\frac{1}{\alpha C_n^Ts_n} + \nu_n} \simeq \alpha C_n^Ts_n\left(1-\nu_n\alpha C_n^Ts_n\right).
	\]
	Since the p.d.f. of $\nu$ is symmetric about zero, we write
	\begin{equation}
		\label{eq:observation}
		o_n \simeq \alpha C_n^Ts_n\left(1+\nu_n\alpha C_n^Ts_n\right).
	\end{equation}
The partially observable system~\eqref{eq:dynamics} and~\eqref{eq:observation} depends on the unknown parameters $\alpha$, $\gamma$, $\sigma_p$ and $\sigma_o$. They are estimated by running several stiffness measurements on test specimens built for that purpose in a calibration phase, described later 
in the experimental part of the paper.

\section{Feedback control law}\label{feedbacklaw}

The below estimator-based control architecture was chosen for its appropriateness to the problem at hand. It relies on the solving of a constrained optimization problem for the whole remaining trajectory at each stage.
\subsection{State estimation}

	The state $s_n$ at stage $n$ is estimated with a Kalman filter consisting of two steps, the state process update and the state measurement update.
   Only the first $n$ entries of the state are considered since the $N-n$ other ones are $0$. Let $\mu_n \in \reals^n$ be the state estimate and $\Sigma_n \in \reals^{n \times n}$ be the state error covariance matrix. We write
	\[ s_n = \mu_n + \delta_n,\]
	where $\delta_n$ is the state estimation error. The observation $o_n$ then reads
\[
\begin{array}{rcl}
		o_n &\simeq& \alpha C_n^Ts_n\left(1+\nu_n\alpha C_n^Ts_n\right)\\
		&=& \alpha C_n^T(\mu_n + \delta_n)\left(1+\nu_n\alpha C_n^T(\mu_n+\delta_n)\right)\\
		&\simeq& \alpha C_n^T(s_n + \mu_nC_n^T\mu_n\nu_n).
		\end{array}
\]
Consistent with standard Kalman filter textbooks~\cite{Anderson79optimalfiltering}, the "state process update" follows from~(\ref{eq:dynamics}) 
	\begin{flalign}
	\begin{aligned}
		\bar{\mu}_{n} &= (\mu_{n-1}^T, a_n)^T,  \\
		\bar{\Sigma}_{n} &= 
		\begin{bmatrix}
		\Sigma_{n-1} & 0 \\
		0 & a_n^4\hat{\sigma}_p^2
		\end{bmatrix},
	\end{aligned}
	\label{postcontrol}
	\end{flalign}
where $a_n = 1/(u_n+\gamma)$. 

The "state measurement update" is
	\begin{flalign}
	\begin{aligned}
		\Sigma_n &= \left(\bar{\Sigma}_n^{-1}+\frac{C_nC_n^T}{\alpha^2\hat{\sigma}_o^2(C_n^T\bar{\mu}_{n})^4}\right)^{-1} \\
		\mu_n &= \Sigma_n\left(\bar{\Sigma}_n^{-1}\bar{\mu}_n + \frac{o_n}{\alpha^3\hat{\sigma}_o^2(C_n^T\bar{\mu}_n)^4}C_n\right).
	\end{aligned}\label{postobs}
	\end{flalign}

    However, measurements are not performed after every control stage. In that case, only the state process update is made, simply followed by
	\begin{flalign}
	\begin{aligned}
		\Sigma_n &= \bar{\Sigma}_n \\
		\mu_n &= \bar{\mu}_n.
	\end{aligned}
	\end{flalign}

Note that this recursive filter is derived from Kalman filtering principles, but it is likely not optimal, in particular because the observation update results from a linearization of the system's output~(\ref{eq:observation}).

	\subsection{Control strategy}\label{costsection}
	At each stage $n$, the control optimizes a cost function, described below. Moreover, the sequence of control inputs must be bounded and decreasing to avoid overhangs, and the predicted stiffness of the finished part must be equal to the desired compliance. The constraints are written
	\begin{equation}
	\label{eq:constraint_1}
		u_{\max} \geq u_n \geq u_{n+1} \geq \dots \geq u_N \geq u_{\min},
	\end{equation}
	\begin{equation}
	\label{eq:constraint_2}
		\alpha\left(\sum_{k=1}^{n-1}c_{n,k}\mu_k+\sum_{k=n}^{N}\frac{c_{n,k}}{u_k + \gamma}\right) = \mathcal{C}^*,
	\end{equation}
	 where $\mathcal{C}^*$ is the target compliance. $u_{\min}$ and $u_{\max}$ are given parameters. 
The cost function is the weighted sum
	\begin{equation}
	\label{eq:costfunction}
	\mathcal{L}_n = \alpha_1\mathcal{L}_{1,n} + \alpha_2\mathcal{L}_{2,n}+ \alpha_3\mathcal{L}_{3,n},
	\end{equation}
where $\alpha_1$, $\alpha_2$ and $\alpha_3$ are positive, and
	\[
	\begin{array}{l}
		\displaystyle \mathcal{L}_{1,n}= \sum_{k=n}^N u_k, \\
		\mathcal{L}_{2,n} = \sum_{k=n-1}^{N-1} (u_{k+1}-u_k)^2 + (u_N-u_{\min})^2, \\
		\displaystyle\mathcal{L}_{3,n} = \alpha^2\hat{\sigma}_p^2\sum_{k=n}^{N} \frac{C_{N,k}^2}{(u_k+\gamma)^4}.
		\end{array}
	\]
The term $\mathcal{L}_{1,n}$ is the quantity of material used. The term $\mathcal{L}_{2,n}$ prevents the series of control inputs to change too abruptly and favors a smooth transition from the initial width of the beam to a final width close to the allowed minimum.
The term $\mathcal{L}_{3,n}$ is the variance of the final compliance of the system after applying the controls $u$ without measurements. It drives the shape of the final beam to regions of the state space where confidence on the final stiffness is higher~\cite{goldshtein2017finite}. 
The cost function~\eqref{eq:costfunction}, the constraints~(\ref{eq:constraint_1}) and~(\ref{eq:constraint_2}),  and the Kalman filter~\eqref{postcontrol} yield an observer-based optimal control strategy: At stage $n$, we write and solve the optimization program
	\begin{equation}
		\begin{array}{rl}
			\mbox{minimize}_{u_n,\dots,u_N} &\mathcal{L} \\
			\mbox{subject to:} &
			u_{\max} \geq u_n \geq u_{n+1} \geq \dots \geq u_N \geq u_{\min}, \\
			&\displaystyle \alpha\left(\sum_{k=1}^{n-1}c_{n,k}\mu_k+\sum_{k=n}^{N}\frac{c_{n,k}}{u_k + \gamma}\right) \leq \mathcal{C}^*.
		\end{array}
	\label{control_law}
	\end{equation}
    In this experiment, the control is applied for a number of stages before performing an observation and recomputing the next controls. This strategy is undoubtedly inspired by published MPC techniques.
    In MPC (see~\cite{mpc} for example), a system is described by the dynamics
    \[
        x(k+1) = f(x(k), u(k)).
    \]
Given $x(k)$ (or its best estimate), the control variable $u(k)$ is determined by minimizing the cost function
    \[
        V= \sum_{i=0}^{L-1} l(\tilde{x}(i), \tilde{u}(i)) + F(\tilde{x}(L))
    \]
over the variables $\tilde{x}(0), \ldots, \tilde{x}(L)$, $\tilde{u}(0), \ldots,\tilde{u}(L-1)$,  under the constraints
  \[
  \tilde{x}(i+1) = f(\tilde{x}(i),\tilde{u}(i)), i = 0, \ldots, L-1, \mbox{ and }\tilde{x}(0) = x(k),
  \]
  where $L$ is the horizon of the prediction, $l(\cdot)$ is the stage cost and $F(\cdot)$ is the terminal cost.
  Denoting optimizing values $\tilde{u}^\ast(0), \ldots,\tilde{u}^\ast(L-1)$, $u(k)$ is given by
  \[
  u(k)= \tilde{u}^{\ast}(0).
\]  
Thus only the first control input $\tilde{u}^{\ast}(0)$ is actually used, and the optimization-based procedure is then repeated at step $k+1$.
\iffalse
    Our control formulation is a finite-time optimal control problem that differs from MPC. Instead of solving an optimization problem on a receding horizon, the whole remaining trajectory is optimized.
    Using MPC on a finite-time problems is justified for real-time systems when the computational cost of optimizing the whole trajectory is excessive, thus the recourse to a smaller fixed prediction horizon. For instance, works that model complex in-layer dynamics such as~\cite{guo2018control} use MPC with a small horizon and could not realistically solve the control problem over the whole trajectory.
    This simple dynamics used in our model make the use of a receding horizon unnecessary.
    %
\fi
Instead, the control strategy used in this paper is to solve the optimal control problem over {\em all} remaining control variables at every remaining stage. This choice is motivated by the fact that the printing process involves a {\em finite} total number of steps and tractable computations. In that regard, our strategy is analogous to the "finite-time" linear quadratic optimal control process~\cite{BrH:75}, although the solution to the optimization problem defined in~(\ref{control_law}) is not explicit and requires solving a convex optimization problem. A detailed discussion discussing this strategy against alternatives is provided below. Note that the initial equality constraint~(\ref{eq:constraint_2}) has been relaxed to a convex inequality constraint. This relaxation is lossless and
    the optimization problem~(\ref{control_law}) yields the same global optimum and optimizers as those that would have been obtained by enforcing the constraint~(\ref{eq:constraint_2}) instead.
The relaxation is also a convex optimization problem~(\ref{control_law}) and it can therefore be solved to optimality at a low computational cost~\cite{BEFB:94, boyd2004convex}. Quantitative computation times are provided below.

\section{Experimental validation}\label{experiments}

\begin{figure}
		\centering
		\framebox{\parbox{.45\textwidth}{\includegraphics[width=.45\textwidth]{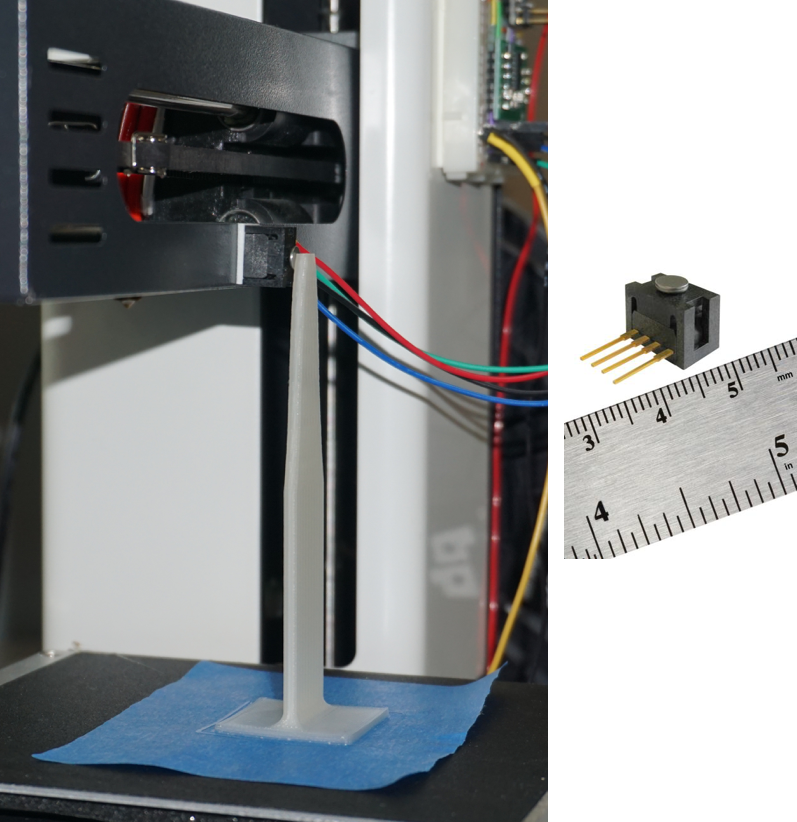}}}
		\caption{Load sensor setup. Right: Individual load sensor. Left, load sensor mounted on 3D printer to measure the force applied on the deflected, partially printed beam.}
		\label{img:base_measurement}
	\end{figure}

The printer used for the closed-loop control experiment is a Monoprice MP Select Mini 3D Printer V2~\cite{monoprice}. The measurement system used here is a Honeywell force sensor mounted directly on the printer's side, see Fig.~\ref{img:base_measurement}. Sensor characteristics are given in~\cite{FSG15N1A}. Displacements are measured by counting the number of steps of the print table stepper motor.
	The goal of the performed experiments is to validate that applying feedback control to the printing process can improve the global properties of the printed object by compensating for model uncertainty and process and sensor noises. 
\subsection{Parameter estimation}
\begin{figure}[!htb]
		\centering
		\framebox{\parbox{.49\textwidth}{\includegraphics[width=.49\textwidth]{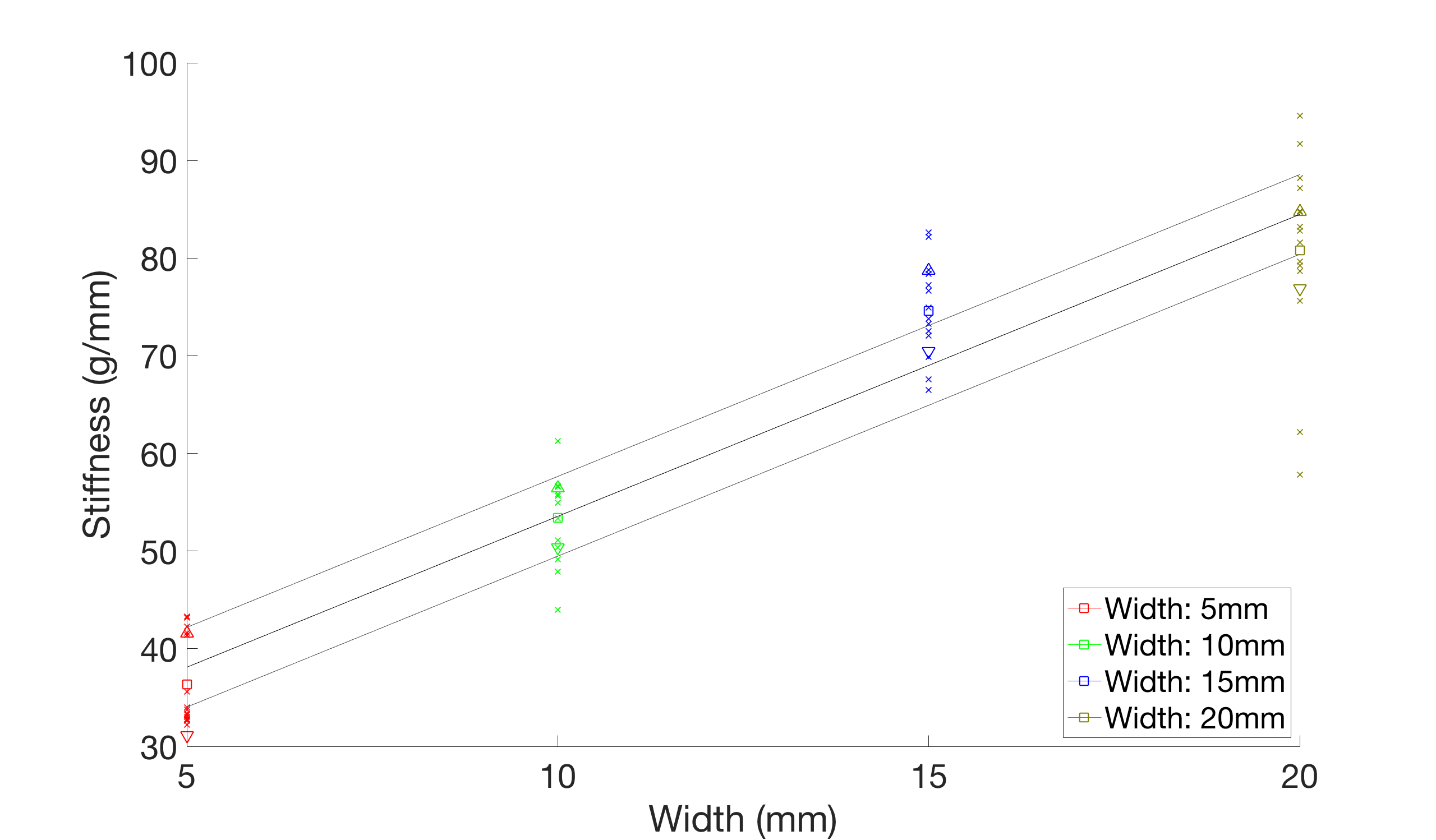}}}
		\caption{Measured stiffness of the twelve test specimens based on five successive measurements per specimen. $\times$ symbols represent individual measurements while their average over specimens with same width is represented by $\square$, and their standard deviation by $\nabla$ and $\Delta$.}
		\label{fig:test_parts_stiff_regression}
	\end{figure}

The parameters $\alpha$, $\gamma$, $\sigma_p$, and $\sigma_o$ introduced in Eqs.~\eqref{eq:secondmom}, \eqref{eq:dynamics}, and~\eqref{eq:observation} are estimated by printing specimens and measuring their compliance.
The chosen specimen height is \SI{50}{\milli\metre} or $N=250$ layers with \SI{0.2}{\milli\metre} individual height, the thickness is \SI{3}{\milli\metre},
while the input width $u$, measured in millimeters, satisfies $u \in \left\{ 5, \;10, \;15 ,\; 20\right\}$. Alternatively, we write $u_i = 5i$, $i = 1,\ldots,4$.
For each width, $p=3$ specimens are printed. The infill density is constant (50\%). For each specimen, $q = 5$ measurements are made after completion (therefore $n=N=250$).
Let $m=4$ be the number of different specimen widths.
A specimen of width $u$ is defined uniquely by the pair $(i,j)$ with $i\leq m$ and $j\leq p$, we consider a stiffness measurement $l$ with $l\leq q$. Combining equations \eqref{eq:dynamics} and \eqref{eq:observation} yields the measured compliance $\mathcal{C}_{i,j,l}$
\begin{align*}
\mathcal{C}_{i,j,l} =&\ \frac{\alpha}{u_i+\gamma}\left(\sum_{k=1}^{n}c_{n,k} + \frac{1}{u_i+\gamma}\sum_{k=1}^{n}c_{n,k}\epsilon_{i,j,k}\right) \\
	&\times \left(1+ \frac{\alpha\nu_{i,j,l}}{u_i+\gamma}\left(\sum_{k=1}^{n}c_{n,k} + \frac{1}{u_i+\gamma}\sum_{k=1}^{n}c_{n,k}\epsilon_{i,j,k}\right)\right),
 	\end{align*}
	where $\epsilon_{i,j,k}$ is the value of the process noise of the layer $k$ of the specimen $(i,j)$ and $\nu_{i,j,l}$ the observation noise of the measurement $l$ taken on the same specimen.
 	Then $\hat{\mathcal{C}}_{i} = \frac1{pq}\sum_{j=1}^{p}\sum_{l=1}^{q}\mathcal{C}_{i,j,l}$ is an unbiased estimator of the expectation of the compliance of a test specimen of given input width $u_i$, since
\[
		\mathbb{E}(\mathcal{C}_i) = \frac{\alpha}{u_i+\gamma}\sum_{k=1}^{n}c_{n,k}, \mbox{ or } \mathbb{E}(\mathcal{K}_i) = (u_i+\gamma )/(\alpha \sum_{k=1}^{n}c_{n,k}),
 	\]\label{compliance_est_1}
where $\mathcal{K}_i$ is the corresponding stiffness and expected values are taken over $\epsilon_{i,j,k}$ and $\nu_{i,j,jl}$.
	Since $u_i$ and $c_{n,k}$ are known, an affine regression on the different estimates $\hat{\mathcal{K}}_{i}$  vs. $u_i$, shown in Fig.~\ref{fig:test_parts_stiff_regression}, gives $\alpha$ and $\gamma$.
 	
%
%
%
 	\iffalse
 	\begin{figure}[!htb]
 		\centering
 		\framebox{\parbox{.49\textwidth}{\includegraphics[width=.49\textwidth]{test_parts_measurements}}}
 		\caption{Series of load and deflection measurements performed on 12 test specimens. Different colors represent different widths.}
 		\label{fig:test_parts_measurements}
 	\end{figure}
 \fi 
Next, consider
  	\begin{align}
 	\label{eq:compliance_est_2}
 	\hat{\mathcal{C}}_{i,j} = \frac1q\sum_{l=1}^{q}\mathcal{C}_{i,j,l}=\frac{\alpha}{u_i+\gamma}\left(\sum_{k=1}^{n}c_{n,k} + \frac{1}{u_i+\gamma}\sum_{k=1}^{n}c_{n,k}\epsilon_{i,j,k}\right).
 	\end{align}
  $\hat{\mathcal{C}}_{i,j}$ is an unbiased estimator of the compliance of the specimen $(i,j)$ and we obtain an estimator $\hat{\sigma}_o^2$ 
 by writing
 	\[
 		\hat{\sigma}_o^2 = \frac{1}{mpq}\sum_{i=1}^m \sum_{j=1}^p \sum_{l=1}^q \left(\frac{\mathcal{C}_{i,j,l}-\hat{\mathcal{C}}_{i,j}}{\hat{\mathcal{C}}_{i,j}^2}\frac{u_i+\gamma}{\alpha}\right)^2.
 	\]
From the estimator \eqref{eq:compliance_est_2}, we can also isolate $\sum_{k=1}^{n}c_{n,k}\epsilon_{i,j,k}$, which is a random variable with a normal distribution of mean $0$ and variance $\sigma_p^2\sum_{k=1}^nc_{n,k}^2$. Thus we can estimate the value of $\sigma_p^2$ by writing
 	\[
 		\hat{\sigma}_p^2 = \left(\sum_{k=1}^{n}c_{n,k}\epsilon_{i,j,k}\right)^2/\sum_{k=1}^nc_{n,k}^2.
 	\]

The values found by the described procedure are given in Table \ref{table:params}.

\begin{table}[h]
	\centering
	\caption{Estimated values of the model parameters}
	\renewcommand{\arraystretch}{1.5}
	\begin{tabular}{|c|c|}
		\hline
		Parameter & Value \\ \hline
		$\alpha$ & \SI{1.035e-8}{\per\gram} \\ \hline
		$\gamma$ & \SI{7.326}{\milli\metre}   \\ \hline
		$\sigma_p$ & \SI{19.064}{\milli\metre} \\ \hline
		$\sigma_o$ &  \SI{3.907}{\milli\metre} \\ \hline
	\end{tabular}
	\label{table:params}
\end{table}
It is well-worth noting that the value of $\sigma_p$ invalidates the approximation made in equation~(\ref{eq:firstorder}). However, the validation of the model is not  central to this paper, whose focus is to experimentally validate the usefulness of feedback control to adjust global stiffness properties, using {\em any} control law regardless of the way it was obtained. The experiment therefore proceeds using the foregoing model-based control law even if the model is used beyond its domain of validity. As shown below, this experiment is successful.

\subsection{Open-loop control experiment}
Several specimens are printed open-loop. The printed beam resembles that shown in the diagram shown in Fig.~\ref{img:shape_diagram}. The total beam height is \SI{10}{\centi\meter} or $N=500$ layers, thickness is \SI{3}{\milli\meter}. A rectangular base of \SI{250}{} layers (\SI{5}{\centi\meter}) with width of \SI{20}{\milli\meter} is printed first. The widths of the remaining \SI{250}{} layers are obtained by solving~(\ref{control_law}) once, with $k = 250$. The resulting set of controls is implemented open-loop. Minimum and maximum layer widths are \SI{5}{} and \SI{20}{\milli\meter}, respectively, and the target compliance is \SI{0.12}{\milli\meter\per\gram}, which is equivalent to a stiffness of approximately \SI{8.333}{\gram\per\milli\meter}. Total print time is about six hours.
Stiffness measurements taken on the five specimens are given in Table~\ref{table:open}.
	\begin{table}[h]
		\centering
		\caption{Measured stiffness and error with the target (\SI{8.333}{\gram\per\milli\meter}) of the finished open-loop control specimens averaged over five measurements}
		\renewcommand{\arraystretch}{1.5}
		\begin{tabular}{|c|c|c|}
			\hline
			& Stiffness (\SI{}{\gram\per\milli\meter}) & Error \\ \hline
			Specimen 1 & 10.3 & 1.962 (23.54\%) \\ \hline
			Specimen 2 & 10.66 & 2.326 (27.91\%) \\ \hline
			Specimen 3 & 13.34 & 5.008 (60.09\%) \\ \hline
			Specimen 4 & 10.84 & 2.503 (30.03\%) \\ \hline
			Specimen 5 & 10.3 & 1.965 (23.58\%) \\ \hline
			Average & 11.09 & 2.753 (33.03\%) \\ \hline
			Standard deviation & 1.282 & --- \\ \hline
		\end{tabular}
		\label{table:open}
	\end{table}

	\subsection{Closed-loop control experiment}
\begin{figure}[!htb]
		\centering
		\framebox{\parbox{.45\textwidth}{\includegraphics[width=.45\textwidth,trim={5cm 0 5cm 0},clip]{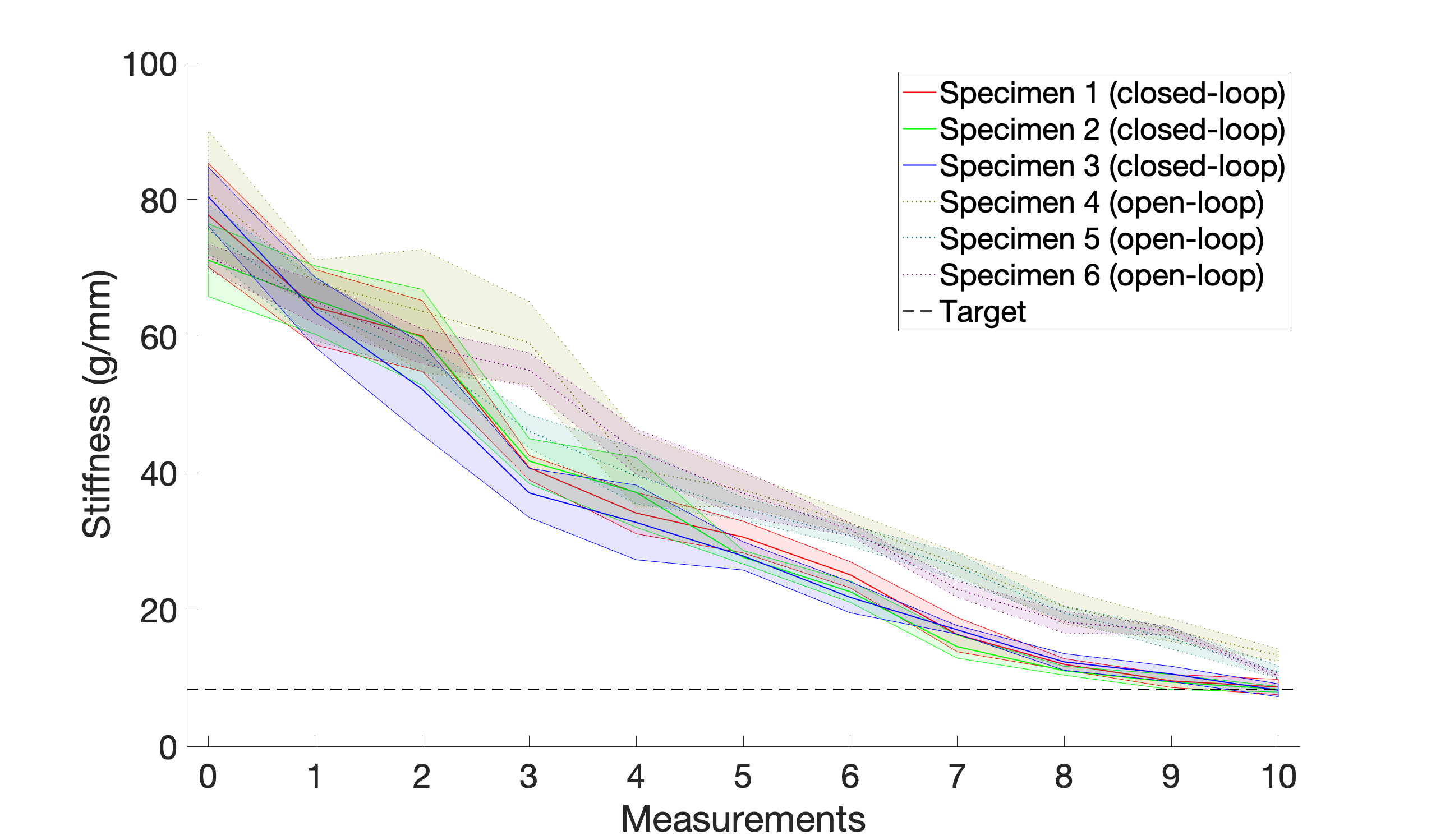}}}
		\caption{Measured stiffness of three closed-loop control specimens versus three open-loop control specimens based on five successive measurements every \SI{25}{} stages. The shaded areas represent the standard deviations of the five measurements. 
		}
		\label{fig:all_stiffness_plot}	
	\end{figure}

Three specimens are then printed using the foregoing closed-loop control strategy and the same target compliance. Five measurements are made every 25 layers and used by the control law given by~(\ref{postcontrol}), (\ref{postobs}), and~(\ref{control_law}). Timing information is as follows:
 \begin{itemize}
	 	\item Printing the part foundation ($\approx$ three hours).
	 	\item Printing under closed-loop control ($\approx$ \si{3.5} hours), decomposed into
		\begin{itemize}
	 		\item measuring stiffness ($\approx$  \si{30} seconds per measurement),
	 		\item computing estimates and control variables ($\approx$  \si{30} seconds per measurement),
	 		\item printing 25 layers ($\approx$  \si{20} minutes per measurement).
	 	\end{itemize}
	 \end{itemize}
Sensing, estimation and control tasks are therefore negligible time-wise, relative to print tasks.
The stiffness measurements made after the parts are printed are shown in Table~\ref{table:closed}. The stiffnesses measured during the printing process are also plotted in Fig.~\ref{fig:all_stiffness_plot}. We also provide stiffness measurements for the last three specimens that were printed open-loop for comparison purposes. %
	\begin{table}[h]
		\centering
		\caption{Measured stiffness of the finished closed-loop control specimens based on five successive measurements and error with the target (\SI{8.333}{\gram\per\milli\meter})}
		\renewcommand{\arraystretch}{1.5}
		\begin{tabular}{|c|c|c|}
			\hline
			& Stiffness (\SI{}{\gram\per\milli\meter}) & Error \\ \hline
			Specimen 1 & 8.683 & 0.3497 (4.197\%) \\ \hline
			Specimen 2 & 8.352 & 0.0189 (0.227\%)\\ \hline
			Specimen 3 & 8.205 & 0.1287 (1.545\%) \\ \hline
			Average & 8.413 & 0.1658 (1.989\%) \\ \hline
			Standard deviation & 0.245 & --- \\ \hline
		\end{tabular}
		\label{table:closed}
	\end{table}

\subsection{Discussion}
We now discuss the impact of feedback control on 3D printed parts from the perspective of their stiffness properties.
	\subsubsection{Model uncertainty}
The average stiffness of the specimens printed open-loop is systematically higher than the target stiffness, most likely because of the
 inaccuracy of the model that is used to derive the optimal controls, and a tendency for this model to underestimate the predicted final stiffness of the printed object. Nevertheless, the closed-loop controller, despite being based on the same inaccurate model, is able to lower the average final stiffness value error from \SI{2.573} {\gram\per\milli\meter} to \SI{0.1658}{\gram\per\milli\meter}, an improvement of almost \SI{94}{\percent}. This result shows the relevance of closed-loop control for handling model inaccuracies.

\subsubsection{Noise rejection}
The standard deviations of the final stiffness of the parts built via closed-loop control is \SI{0.245}{\gram\per\milli\meter}, an 80\% improvement over the \SI{1.282}{\gram\per\milli\meter} standard deviation observed on parts printed open-loop. %

 It would be possible to adjust the open-loop controller based on the result of a first print to result in a next printed specimen having a final stiffness closer to the specifiec requirement.
 Similar approaches exist in rapid thermal processing for the manufacturing of silicon wafers, for which sensor architectures are designed to monitor the manufacturing process in order to improve it subsequently~\cite{freed2001autonomous}. However, this approach would not be able to handle process variability and external disturbances perturbing the printing process.
This variability is "process noise" and cannot be improved by a more accurate printing model, thus fully validating the use of feedback control in the context discussed in this paper.
Along the lines of that comparison with rapid thermal processing, our approach is more akin to the in-process monitoring and control of the manufacturing process~\cite{norman1991improvement, cho1994contribution}.

\section{Suggestions for further research}\label{suggestions}

	\subsection{Control algorithms}
    The control law and model used in this paper are not optimally designed or tuned. Several improvements are probably possible, including 
 reducing computational complexity or using a better model of the printing process, potentially accounting for layer-to-layer and in-layer dynamics as in~\cite{sammons2019two}. A more accurate model is likely to improve the results of model-based control policies, but would possibly increase computational complexity. Indeed, as mentioned, solving the optimization problem~(\ref{control_law}) is easy because the identified model makes all constraints convex and the number of variables relatively low. If using a more accurate model yields expensive computations, an alternate "plan and regulate" strategy could be adopted, whereby an optimal sequence of inputs (layer widths), together with an optimal reference stiffness profile could be computed only once prior to printing operations. Given such a stiffness profile, low-complexity trajectory-following techniques, such as PID~\cite{sammons2018repetitive} or receding-horizon control over a few steps~\cite{zomorodi2016extrusion,guo2017distributed} could then be used to regulate the "stiffness trajectory" against the pre-computed reference trajectory. It must be noted, however, that PID control may sometimes generate undesirable corrective actions: For example, PID may create accidental "overhangs".
 		
	\subsection{Multiple structural and geometrical requirements}
		Another direction of research in the lines of this work is to study the implications of having multiple requirements on the structural and geometrical properties of the part being built. Since our experiment was restricted to one dimension, adding stiffness requirements in several directions for a 3-dimensional part would be a natural extension. Instead of controlling the width of the part being built, a different parameterization of each layer shape would be required. A possible parameterization could come from finite-element analysis, with an initial model made of finite elements satisfying the given requirements being adapted as the part is being built. 
		
	\subsection{Sensing}
		As the complexity of the constraints increases, it is clear that more sensing capabilities are required. Finding efficient non-destructive methods to measure a part's structural properties while it is being built is also a main direction for future research. Moreover, if sensing requires to pause the printing process like in this experiment, the process of deciding when to make measurements could be optimized with respect to the precision required on the structural requirements. More stringent requirements would incur more frequent measurements. It is also possible to imagine a mechanism where measurements would not be made at fixed intervals, but would be triggered on-line as the incertitude increases too much, for instance when the last measurement give a result too dissimilar to what was expected or when the last control steps have particularly uncertain outcomes.

\section{Conclusion}\label{conclusion}

	This paper presents an experiment that validates using feedback control for 3D printed parts to meet global stiffness properties accurately. A simple dynamical model of the manufacturing process of a cantilever beam was calibrated from printed test specimens and used to derive an observer-based  optimal control law.  All measurements were performed \emph{in situ} by pausing the printing process and using the printer itself to measure the compliance of the cantilever beam during printing. A comparative study of specimens printed open-loop and closed-loop validates the hypothesis that feedback control improves printing performance from the viewpoint of managing model uncertainties and rejecting process and sensor noise for the application of interest in this paper.

\addtolength{\textheight}{-4.5cm}   %

\section*{ACKNOWLEDGMENT}\label{acknowledgment}

This research was supported in part by NSF awards CNS \#1446758 and CNS \#1544332. The authors would like to thank the reviewers of this paper for their precious comments, suggestions, and ideas for further research. The authors would also like to extend their warm thanks to Kerianne Hobbs and Dan Berrigan from the Air Force Research Laboratory for introducing the authors to additive manufacturing and pointing out the role feedback control could play in this field.

\bibliography{biblio/kevin}

\begin{thebibliography}{10}
\providecommand{\url}[1]{#1}
\csname url@samestyle\endcsname
\providecommand{\newblock}{\relax}
\providecommand{\bibinfo}[2]{#2}
\providecommand{\BIBentrySTDinterwordspacing}{\spaceskip=0pt\relax}
\providecommand{\BIBentryALTinterwordstretchfactor}{4}
\providecommand{\BIBentryALTinterwordspacing}{\spaceskip=\fontdimen2\font plus
\BIBentryALTinterwordstretchfactor\fontdimen3\font minus
  \fontdimen4\font\relax}
\providecommand{\BIBforeignlanguage}[2]{{%
\expandafter\ifx\csname l@#1\endcsname\relax
\typeout{** WARNING: IEEEtran.bst: No hyphenation pattern has been}%
\typeout{** loaded for the language `#1'. Using the pattern for}%
\typeout{** the default language instead.}%
\else
\language=\csname l@#1\endcsname
\fi
#2}}
\providecommand{\BIBdecl}{\relax}
\BIBdecl

\bibitem{parthasarathy2011design}
J.~Parthasarathy, B.~Starly, and S.~Raman, ``A design for the additive
  manufacture of functionally graded porous structures with tailored mechanical
  properties for biomedical applications,'' \emph{Journal of Manufacturing
  Processes}, vol.~13, no.~2, pp. 160--170, 2011.

\bibitem{bourell2009brief}
D.~L. Bourell, J.~J. Beaman, M.~C. Leu, and D.~W. Rosen, ``A brief history of
  additive manufacturing and the 2009 roadmap for additive manufacturing:
  looking back and looking ahead,'' \emph{Proceedings of RapidTech}, pp.
  24--25, 2009.

\bibitem{frazier2014metal}
W.~E. Frazier, ``Metal additive manufacturing: a review,'' \emph{Journal of
  Materials Engineering and Performance}, vol.~23, no.~6, pp. 1917--1928, 2014.

\bibitem{huang2015additive}
Y.~Huang, M.~C. Leu, J.~Mazumder, and A.~Donmez, ``Additive manufacturing:
  current state, future potential, gaps and needs, and recommendations,''
  \emph{Journal of Manufacturing Science and Engineering}, vol. 137, no.~1, p.
  014001, 2015.

\bibitem{mazumder1997direct}
J.~Mazumder, J.~Choi, K.~Nagarathnam, J.~Koch, and D.~Hetzner, ``The direct
  metal deposition of {H13} tool steel for {3-D} components,'' \emph{JOM},
  vol.~49, no.~5, pp. 55--60, 1997.

\bibitem{el2008phase}
H.~El~Kadiri, L.~Wang, M.~F. Horstemeyer, R.~S. Yassar, J.~T. Berry,
  S.~Felicelli, and P.~T. Wang, ``Phase transformations in low-alloy steel
  laser deposits,'' \emph{Materials Science and Engineering: A}, vol. 494,
  no.~1, pp. 10--20, 2008.

\bibitem{thijs2010study}
L.~Thijs, F.~Verhaeghe, T.~Craeghs, J.~Van~Humbeeck, and J.-P. Kruth, ``A study
  of the microstructural evolution during selective laser melting of
  {Ti--6Al--4V},'' \emph{Acta Materialia}, vol.~58, no.~9, pp. 3303--3312,
  2010.

\bibitem{zheng2008thermal}
B.~Zheng, Y.~Zhou, J.~E. Smugeresky, J.~M. Schoenung, and E.~J. Lavernia,
  ``Thermal behavior and microstructural evolution during laser deposition with
  laser-engineered net shaping: {Part I. Numerical calculations},''
  \emph{Metallurgical and materials transactions A}, vol.~39, no.~9, pp.
  2228--2236, 2008.

\bibitem{bontha2009effects}
S.~Bontha, N.~W. Klingbeil, P.~A. Kobryn, and H.~L. Fraser, ``Effects of
  process variables and size-scale on solidification microstructure in
  beam-based fabrication of bulky {3D} structures,'' \emph{Materials Science
  and Engineering: A}, vol. 513, pp. 311--318, 2009.

\bibitem{antonysamy2012microstructure}
A.~A. Antonysamy, ``Microstructure, texture and mechanical property evolution
  during additive manufacturing of {Ti6Al4V} alloy for aerospace
  applications,'' 2012.

\bibitem{ahn2002anisotropic}
S.-H. Ahn, M.~Montero, D.~Odell, S.~Roundy, and P.~K. Wright, ``Anisotropic
  material properties of fused deposition modeling {ABS},'' \emph{Rapid
  prototyping journal}, vol.~8, no.~4, pp. 248--257, 2002.

\bibitem{bagsik2010fdm}
A.~Bagsik, V.~Sch{\"o}ppner, and E.~Klemp, ``Fdm part quality manufactured with
  ultem* 9085,'' in \emph{Polymeric materials 2010}.

\bibitem{sammons2019two}
P.~M. Sammons, D.~A. Bristow, and R.~G. Landers, ``Two-dimensional modeling and
  system identification of the laser metal deposition process,'' \emph{Journal
  of Dynamic Systems, Measurement, and Control}, vol. 141, no.~2, p. 021012,
  2019.

\bibitem{guo2018control}
Y.~Guo, J.~Peters, T.~Oomen, and S.~Mishra, ``Control-oriented models for
  ink-jet 3d printing,'' \emph{Mechatronics}, vol.~56, pp. 211--219, 2018.

\bibitem{tang2011layer}
L.~Tang and R.~G. Landers, ``Layer-to-layer height control for laser metal
  deposition process,'' \emph{Journal of Manufacturing Science and
  Engineering}, vol. 133, no.~2, p. 021009, 2011.

\bibitem{lu2014layer}
L.~Lu, J.~Zheng, and S.~Mishra, ``A layer-to-layer model and feedback control
  of ink-jet 3-d printing,'' \emph{IEEE/ASME Transactions on Mechatronics},
  vol.~20, no.~3, pp. 1056--1068, 2014.

\bibitem{mpc}
D.~Mayne, ``Nonlinear model predictive control: Challenges and opportunities,''
  in \emph{Nonlinear model predictive control}.\hskip 1em plus 0.5em minus
  0.4em\relax Springer, 2000, pp. 23--44.

\bibitem{guo2017distributed}
Y.~Guo, J.~Peters, T.~Oomen, and S.~Mishra, ``Distributed model predictive
  control for ink-jet 3d printing,'' in \emph{2017 IEEE International
  Conference on Advanced Intelligent Mechatronics (AIM)}.\hskip 1em plus 0.5em
  minus 0.4em\relax IEEE, 2017, pp. 436--441.

\bibitem{xiong2016closed}
J.~Xiong, Z.~Yin, and W.~Zhang, ``Closed-loop control of variable layer width
  for thin-walled parts in wire and arc additive manufacturing,'' \emph{Journal
  of Materials Processing Technology}, vol. 233, pp. 100--106, 2016.

\bibitem{zomorodi2016extrusion}
H.~Zomorodi and R.~G. Landers, ``Extrusion based additive manufacturing using
  explicit model predictive control,'' in \emph{2016 American Control
  Conference (ACC)}.\hskip 1em plus 0.5em minus 0.4em\relax IEEE, 2016, pp.
  1747--1752.

\bibitem{sammons2018repetitive}
P.~M. Sammons, M.~L. Gegel, D.~A. Bristow, and R.~G. Landers, ``Repetitive
  process control of additive manufacturing with application to laser metal
  deposition,'' \emph{IEEE Transactions on Control Systems Technology}, no.~99,
  pp. 1--10, 2018.

\bibitem{nassar2015intra}
A.~R. Nassar, J.~S. Keist, E.~W. Reutzel, and T.~J. Spurgeon, ``Intra-layer
  closed-loop control of build plan during directed energy additive
  manufacturing of {Ti--6Al--4V},'' \emph{Additive Manufacturing}, vol.~6, pp.
  39--52, 2015.

\bibitem{garanger20183d}
K.~Garanger, T.~Khamvilai, and E.~Feron, ``{3D printing of a leaf spring: A
  demonstration of closed-loop control in additive manufacturing},'' \emph{2nd
  IEEE Conference on Control Technology and Applications}, 2018.

\bibitem{gere1997mechanics}
J.~M. Gere and S.~P. Timoshenko, \emph{Mechanics of materials}, 4th~ed., ser.
  General Engineering Series.\hskip 1em plus 0.5em minus 0.4em\relax PWS Pub
  Co., 1997.

\bibitem{Anderson79optimalfiltering}
B.~D.~O. Anderson and J.~B. Moore, \emph{Optimal Filtering}.\hskip 1em plus
  0.5em minus 0.4em\relax Prentice-Hall, 1979.

\bibitem{goldshtein2017finite}
M.~Goldshtein and P.~Tsiotras, ``Finite-horizon covariance control of linear
  time-varying systems,'' in \emph{Decision and Control (CDC), 2017 IEEE 56th
  Annual Conference on}.\hskip 1em plus 0.5em minus 0.4em\relax IEEE, 2017, pp.
  3606--3611.

\bibitem{BrH:75}
A.~E. Bryson and Y.~C. Ho, \emph{Applied Optimal Control}.\hskip 1em plus 0.5em
  minus 0.4em\relax New York: Hemisphere Publishing, 1975.

\bibitem{BEFB:94}
S.~Boyd, L.~{El~{G}haoui}, E.~Feron, and V.~Balakrishnan, \emph{Linear Matrix
  Inequalities in System and Control Theory}, ser. Studies in Applied
  Mathematics.\hskip 1em plus 0.5em minus 0.4em\relax Philadelphia, PA: {SIAM},
  Jun. 1994, vol.~15.

\bibitem{boyd2004convex}
S.~Boyd and L.~Vandenberghe, \emph{Convex optimization}.\hskip 1em plus 0.5em
  minus 0.4em\relax Cambridge university press, 2004.

\bibitem{monoprice}
\BIBentryALTinterwordspacing
{Monoprice, Inc.} {Monoprice MP Select Mini 3D Printer V2, White}.
  \url{https://www.monoprice.com/Product?p_id=15365}. Accessed 2019-06-09.
  [Online]. Available: \url{https://www.monoprice.com/Product?p_id=15365}
\BIBentrySTDinterwordspacing

\bibitem{FSG15N1A}
\BIBentryALTinterwordspacing
{Honeywell International Inc.} {Honeywell FSG15N1A}.
  \url{https://sensing.honeywell.com/honeywell-sensing-force-sensors-fsg-product-sheet-008028-2-en.pdf}.
  Accessed 2019-06-09. [Online]. Available:
  \url{https://sensing.honeywell.com/honeywell-sensing-force-sensors-fsg-product-sheet-008028-2-en.pdf}
\BIBentrySTDinterwordspacing

\bibitem{freed2001autonomous}
M.~Freed, M.~Kruger, C.~J. Spanos, and K.~Poolla, ``Autonomous on-wafer sensors
  for process modeling, diagnosis, and control,'' \emph{IEEE transactions on
  semiconductor manufacturing}, vol.~14, no.~3, pp. 255--264, 2001.

\bibitem{norman1991improvement}
S.~Norman, C.~Schaper, and S.~Boyd, ``Improvement of temperature uniformity in
  rapid thermal processing systems using multivariable control,'' \emph{MRS
  Online Proceedings Library Archive}, vol. 224, 1991.

\bibitem{cho1994contribution}
Y.~Cho, A.~Paulraj, T.~Kailath, and G.~Xu, ``A contribution to optimal lamp
  design in rapid thermal processing,'' \emph{IEEE transactions on
  semiconductor manufacturing}, vol.~7, no.~1, pp. 34--41, 1994.

\end{thebibliography}

\end{document}